\documentclass[10pt, reprint, aps, amsmath,amssymb,twocolumn,article]{revtex4-1}

\usepackage{epigraph}
\usepackage{graphicx}
\usepackage{dcolumn}
\usepackage{algorithm,algpseudocode}
\usepackage{bm}
\usepackage[font=scriptsize]{caption}
\usepackage{booktabs}
\usepackage{subcaption,graphicx}
\algnewcommand{\To}{\textbf{To }}
\algnewcommand\Input{\item[\textbf{Input:}]}%
\algnewcommand\Output{\item[\textbf{Output:}]}%

\begin{document}

\title{Linearity Analysis of the Common Collector Amplifier, or Emitter Follower}

\author{Luciano da F. Costa}
\email{luciano@ifsc.usp.br, copyright LdaFcosta}
\affiliation{S\~ao Carlos Institute of Physics, IFSC-USP,  S\~ao~Carlos, SP,~Brazil}

\date{\today}

\begin{abstract}
A recently introduced Early modeling of transistors is applied to the study of the common collector amplifier
(or emitter follower), an important type of electronic circuit typically employed as buffer, being characterized by near unit voltage
gain, high input resistance, and low output resistance.  The Early equivalent model is applied to derive a
simple representation that is simple and yet capable of incorporating the transistor non-linearities implied by
the Early effect.  Mathematical expressions are obtained describing completely the circuit operation in terms
of currents and voltages, allowing accurate estimation of the average voltage gain, total harmonic distortion (THD),
and average input and output resistances.  Prototypes of small signal silicon transistors of types NPN and PNP
obtained in a previous work are used to discuss the respectively implied properties of the common collector transistor.
In addition to confirming the importance of the trade-off of current gain for the desired properties, it is
also shown that sub-optimal performance can be obtained in case the base and emitter resistances are not properly
chosen.  Even so, the limited current gain implied by real-world NPN and PNP small signal silicon devices implies
some performance constraints.  In particular, it has been observed that the THD tends to be larger for PNP devices
than NPN counterparts with the same average current gain.  The obtained results pave the way not only to 
complementary analytical studies, but also provide guidance for design and implementation of improved common collector
configurations.
\end{abstract}

\maketitle

\setlength{\epigraphwidth}{.49\textwidth}
\epigraph{``Lo specchio ora accresce il valore alle cose, ora lo nega.''}{Le citt\`a invisibili, I. Calvino}

\section{Introduction}

One of the main concerns in analog electronics regards the \emph{linearity} of the amplification and other related operations
such as filtering.  Given that the cornerstone of linear electronics, namely transistors, are intrinsically non-linear, it becomes
particularly important to develop and apply methods capable of quantifying the effects of these non-linearities on the circuit
operation.  This can be done by developing models, and respective equivalent circuits, of the employed transistors, and using
them to derive mathematical expressions describing the circuit and respective operation.  There are two main such approaches 
to transistor and circuit modeling: (1) simple, linear models such as those based on the fixed current gain $\beta$ and output
resistance $R_o$ (e.g.~\cite{roehr:1963,gray:1969,gray:1990,thomson:1976,jaeger:1997}), and (2) relatively sophisticated models 
such as the Gummel-Poon and related approaches (e.g.~\cite{gummel:1970,moinian:2000}), capable of
incorporating several effects into the transistor model, but requiring substantial efforts for the respective solution.  As a matter
of fact, the latter type of approaches has been typically performed through computational simulations, being rarely used analytically for the
study of circuit configurations.  

Recently, a new approach to transistor modeling has been reported (e.g.~\cite{costaearly:2017,costaearly:2018,costaequiv:2018}) 
that provides simplicity at the level of the
simplest modeling with fixed $\beta$ and $R_o$, but still allowing considerable accuracy in representing the transistor non-linearities
as implied by the fanned, converging characteristic isolines implied by the Early effect (e.g.~\cite{early:1952,jaeger:1997}).  
This approach, henceforth
called Early modeling, is characterized by using only two parameters to represent a given transistor: its estimated, fixed Early voltage
$V_a$, as well as a proportionality parameter $s$.  Actually, it was the introduction of the latter 
parameter~\cite{costaearly:2017,costaearly:2018,costaequiv:2018} that paved the way to a complete Early approach to transistor
modeling, after being
experimentally verified for hundreds of small signal silicon and germanium junction transistors that the characteristic isolines 
angle $\theta$ varies linearly with the base current $I_B$, i.e $\theta = s I_B$.  It follows that the collector output resistance becomes
$R_o = 1/tan(s I_B) \approx 1/(sI_B)$.   The equivalent circuit of the Early approach therefore corresponds to a very simple configuration
involving a fixed voltage source $V_a$ in series with the $I_B$-modulated resistance $R_o$.  As such, the respectively derived
representation of several types of circuits allow mathematical expressions to be obtained describing to good completeness and
accuracy the respective operation.

Regarding the experimental estimation of $V_a$ and $s$ from real-world transistors, as required for their respective studies, 
a simple and yet accurate approach has been reported~\cite{costaearly:2018} that involves a Hough transform accumulation 
scheme (e.g.?\cite{shapebook,hough:1962,}) for the identification of the intersection of the transistors
isolines (obtained by linear regression) defining the Early voltage $V_a$.  The proportionality parameter $s$ can then be obtained 
by least minimum squares from the relationship between $\theta$ and $I_B$.  The overall estimation of $V_a$ and $s$ of a given
transistor takes less than 2 min by using the author's acquisition and analysis system, and can be substantially speeded up by
using more powerful equipment.  So, from the analytical point of view, the Early approach to characterizing and modeling transistor 
results simple and yet  capable of representing the isolines non-linearities.  Also from the experimental point of view, the reported
procedures for estimating $V_a$ and $s$ can be easily and conveniently accomplished by using relatively simple and inexpensive
resources.  These features make of the Early modeling approach a viable alternative for analyzing several types of transistors in the
most diverse circuit configurations.

The simplicity of the Early approach, allied to its ability to represent 
the transistor non-linearities implied by the Early effect, have paved the way to a number of successful respective applications to
the analysis and better understanding of several electronics effects and circuits, including stability with power 
oscillations~\cite{costaearly:2018}, study of complementary geometry transistors in push-pull amplifiers~\cite{costaearly:2018},
the effect of reactive loads on amplification~\cite{costareact:2018}, as well as the study of the common emitter amplifier 
considering the transistor non-linearities~\cite{commemitt:2018}.

In the present work, we approach the frequently used circuit configuration, shown in Figure~\ref{fig:follower}, known as the \emph{common
collector amplifier} (e.g.~\cite{roehr:1963,gray:1990,thomson:1976,jaeger:1997}).  
The distinguishing features of this circuit include: (i) near unit voltage gain; (ii) large input resistance;
ad (iii) small output resistance.   As such, this circuit is frequently employed as a \emph{buffer}.  Yet, most of the analysis of this
circuit, such as those using the hybrid-pi model, assumes the transistor current gain $\beta$ not to vary with either the collector
voltage $V_C$ or the collector current $I_C$.  Neither of this is observed in real-world circuits, where the Early effect implied
non-linearities conspire to add distortion to the amplified signal.  In practice, the relatively high level of negative feedback accounted
by $R_E$ has been assumed to be enough to eliminate, or at least reduce considerably, such distortions.   

However, it remains to be better understood to which level this negative feedback can cope with the transistor non-linearities.  Here, 
we use the recently introduced Early methodology in order to study the common collector amplifier while taking into account the transistor 
non-linearities.  This is done by using the Early equivalent model to derive a respective circuit representation of this amplifier in 
terms of the two Early parameters $V_a$ and $s$.  Because of the simplicity of the Early equivalent model, it becomes possible 
to derive mathematical expressions describing to a good level the operation of the common emitter circuit in presence of transistor 
non-linearities.  This approach allows an interesting compromise between linear models of transistors and the more complete
and demanding Gummel-Poon aproach.

Several interesting results are obtained, including the verification of the fact that the level of negative feedback
typically adopted cannot completely eliminate the transistor non-linearities, implying in substantial remaining total harmonic
distortion (THD).  While transistors with larger average current gains allow better buffering properties, it is verified that
NPN transistors yield substantially smaller THD than PNP counterparts with the same average current gain.   

This work is organized as follows.  It starts by presenting the common collector configuration, and then applies the Early
equivalent model in order to obtain a respective overall circuit.  Mathematical expressions describing the operation of this
circuit in terms of the involved currents and voltages are analytically obtained, and then used to study the effect of the
transistor non-linearity, given certain levels of negative feedback, over the circuit operation and implied distortions.  The
work concludes with suggestions for further research.

\section{The Common Collector Amplifier}

\begin{figure}[h!]
\centering{
\includegraphics[width=7cm]{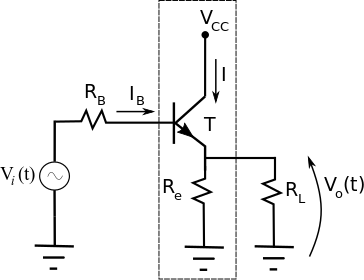}
\caption{The common collector amplifier configuration considered in the present work.  The portion of the circuit within the
dashed box represents what is commonly called the common collector amplifier, but we consider a more complete circuit
incorporating a base resistance $R_B$ (which can also be understood as input source resistance) and the load resistance $R_L$.
The emitter resistance $R_e$ is responsible for implementing negative feedback.   The input voltage
source incorporates a DC level, i.e.~$V_i(t) = A \left[ sin(2 \pi f_o t) + 1 \right]$.}
\label{fig:follower}}
\end{figure}

Figure~\ref{fig:follower} depicts the common collector amplifier as adopted in this work.  The common collector proper 
commonly refers to the portion of the circuit within the dashed box.  For generality's sake, we incorporate a base 
(or input source) resistance $R_B$ as well as the load resistance $R_L$.  Negative feedback is provided by the emitter 
resistance $R_e$, as well as by the source resistance $R_L$ when this is attached to the circuit output.  For simplicity's 
sake, we assume that the input voltage source includes a DC bias, i.e.~$V_i(t) = A \left[ sin(2 \pi f_o t) + 1 \right]$.  
Observe that it is also possible to adapt the following developments to the AC situation in which both the input source and output 
resistance are attached to the circuit by using decoupling capacitors.

\begin{figure}[h!]
\centering{
\includegraphics[width=9cm]{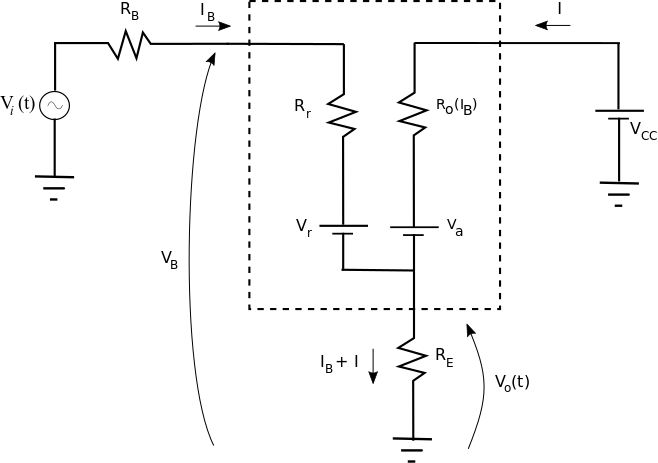}
\caption{The common emitter circuit representation obtained by using the Early equivalent circuit of the NPN transistor
(within the dashed box). Observe that $R_E$ is not necessarily equal to $R_e$.  
Also, recall that $R_o(I_B) = 1/tan(s I_B) \approx 1/(s I_B)$.  $V_a$ is the Early voltage and $s$ is the proportionality parameter.
The input mesh is modeled as usual, i.e.~as a diode with resistance $R_r$ and offset voltage $V_r$.}
\label{fig:follower_equiv}}
\end{figure}

Now, it is possible to use the
Early equivalent circuit~\cite{costaequiv:2018} in order to derive, from the common collector amplifier circuit configuration
in Figure~\ref{fig:follower}, the circuit shown in Figure~\ref{fig:follower_equiv}.

We have that the total current through the emitter resistance is equal to $I_E = I_B + I$.   
Observe that we use a generic emitter resistance $R_E$ that is not necessarily equal to $R_e$.  Also, recall that 
$R_o(I_B) = 1/tan(s I_B) \approx 1/(s I_B)$.   By applying Kirchhoff's current and voltage laws as well as Ohm's law
to this circuit, we obtain the following system of equations:

\begin{eqnarray}
\left\{ 
\begin{array}{c}
   V_i - (R_B + R_r) I_B - V_r - R_E (I + I_B) = 0 \\ 
   V_{CC} -  I/(s I_B) - V_a -  R_E (I +I_B) = 0   
\end{array}  \label{eq:system}
\right.
\end{eqnarray}

The solution of this system, in terms of $I(R_E)$ and $I_B(R_E)$ is given by Equations~\ref{eq:IB} and~\ref{eq:I}, respectively.
The output voltage $V_o(t,R_E)$ can be easily derived from these two currents as given in Equation~\ref{eq:Vo}.

\begin{table*}[h!]
\begin{eqnarray}  \label{eq:sol}
  \gimel(R_E) = \sqrt{\left( 4 s R_1 R_E V_1 + (R_1+R_E+ s R_E (V_2-V_1))   \right)^2}  \\
  I_B(R_E) =   \frac{s R_E (V_1-V_2) - R_1 - R_E + \gimel }{2 s R_1 R_E}  \label{eq:IB} \\
  I(R_E) = \frac{R_1^2 + R_E^2(1 + s (V_2-V_1)) + R_1 R_E(2+s(V_1+V_2))-(R_1+R_E) \gimel}{2 s R_1 R_E^2}  \label{eq:I} \\
  V_1 = V_i(t) - Vr \\
  V_2 = V_{CC} - V_a \\
  V_o(t,R_E) = (I(R_E)+I_B(R_E)) R_E  \label{eq:Vo}
\end{eqnarray}
\end{table*}

We are now ready to obtain the voltage gain $A_v$, which is given in Equation~\ref{eq:Av}.  It is assumed
that $R_E = R_e + R_L$.

\begin{equation}
  A_v = \frac{V_o(t)-min(V_o(t))}{V_i(t)-min(V_i(t))}   \label{eq:Av}  
\end{equation}

The input resistance seen by the input voltage source $V_i(t)$ can be calculated by applying the Thevenin
equivalent theorem applied to the input mesh of the circuit in Figure~\ref{fig:follower_equiv}.  We make
$R_E = R_e + R_L$.  More specifically,
we obtain the Thevenin's equivalent voltage source $V_{Th,in}$ by disconnecting $V_i(t)$.  Under these circumstances, 
there is no base current, i.e.~$I_B=0$, so that $R_o \rightarrow \infty$ and the output mesh is isolated from
$V_CC$, implying $I = 0$.  So, $V_{Th,in} = V_r$.  Now, the input resistance $R_{in}$ seen by $V_i(t)$ can be obtained as:

\begin{equation}
  R_{in} = \frac{V_i - V_r }{I_B(R_E)}   \label{eq:Rin} 
\end{equation}

So, the input resistance is a function of the input voltage $V_i(t)$ and of $I_B$, the latter depending on the Early
parameters $V_a$ and $s$, $R_E$, as well as $V_{CC}$.

In order to derive the output resistance, we disconnect the load resistance from the circuit, implying $R_E = R_e$.
This corresponds to the situation in which the load is open circuited, in which case the output voltage can be obtained
by using Equation~\ref{eq:Vo} with $R_E = R_e$.  Thus, the open circuit voltage is $V_o(R_E = R_e)$.   In order to
obtain the short circuit current, we make $R_E = 0$, so that we have the respective short circuit versions of the
$I_B$ and $I$ currents given as in Equations~\ref{eq:IBs} and~\ref{eq:Is}.  Now, the output resistance $R_{out}$
can be obtained by dividing the open circuit output voltage $V_o(R_E = R_e)$ by the sum of the short circuit output
currents, as expressed in Equation~\ref{eq:Rout}.

\begin{eqnarray}
  I_{Bs} = \frac{V_i - V_r}{R_r + R_B}  \label{eq:IBs} \\
  I_{Es} = \frac{ V_{CC} - V_a}{R_o(I_B)} = \left( V_{CC} - V_a  \right) s I_B  \label{eq:Is}  \\
  R_{out} = \frac{V_{o}(R_E = R_e)}{I_{Bs}+I_{Es}}  \label{eq:Rout}
\end{eqnarray}

So, by using the Early modeling approach, it has been possible to derive analytical expressions giving the 
voltage gain $A_v$ and THD, as well as the input and output resistances $R_{in}$ and $R_{out}$, respectively.
In the next section we explore the behavior of these important parameters with respect to several circuit
and transistor configurations.  The THD values were obtained numerically by using the fast Fourier 
transform~\cite{shapebook,brigham:1988}.

\section{Circuit Analysis}

First, we investigate the effect of $R_B$ and $R_E$ on the circuit properties.  We do this with respect to two
typical transistor configurations found respectively in NPN and PNP silicon bipolar junction transistors~\cite{costaearly:2018}.
These two configurations are $(V_{a,NPN} = -100V, s_{NPN} = 2.5 V^{-1})$ and $(V_{a,PNP} = -50V, s_{PNP} = 5.0 V^{-1})$, which
imply in comparable average current gains $\langle \beta \rangle \approx 250$.    We have adopted $R_r = 30 \Omega$,
$V_r = 0.6V$, and $V_i(t) = 4.0 \left[ sin(2 \pi 200 t) + 1 \right]$.  In all cases in this work, we have $R_e = R_L = 0.5 R_E$.

\begin{figure*}[h!]
\centering{
\includegraphics[width=8cm]{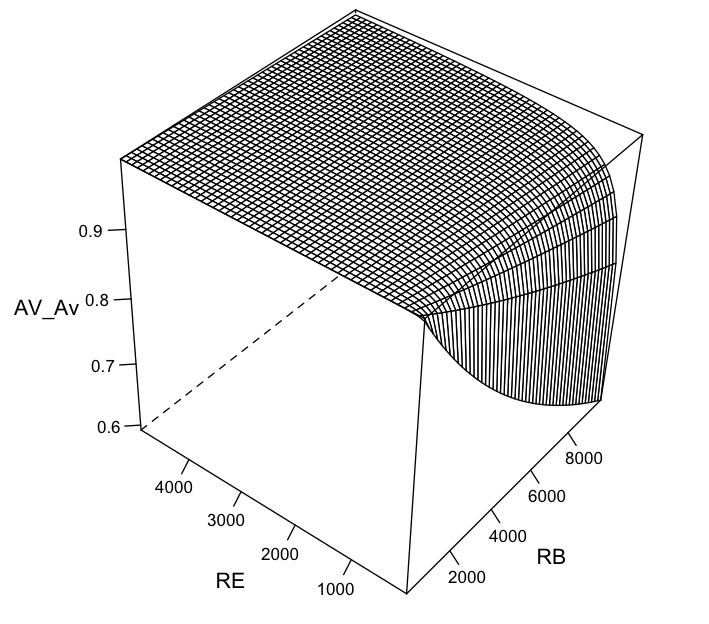}
\includegraphics[width=8cm]{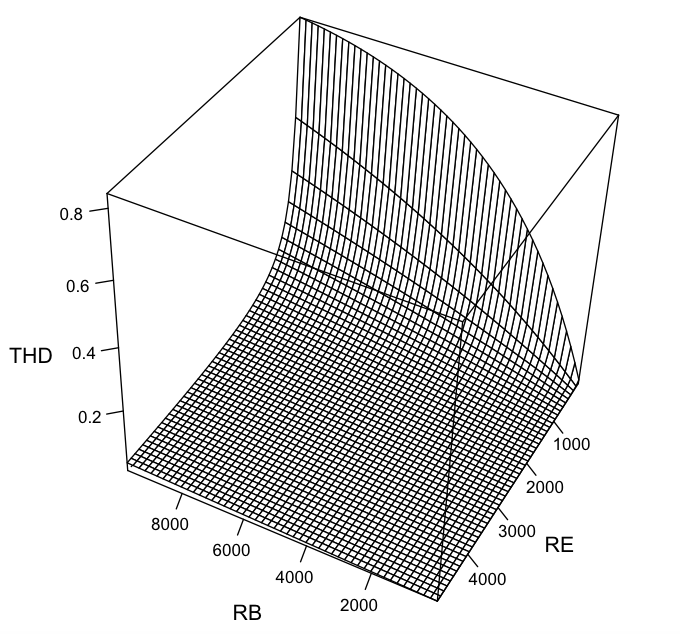}   \\
(a)     \hspace{7.5cm}     (b)   \\
\includegraphics[width=8cm]{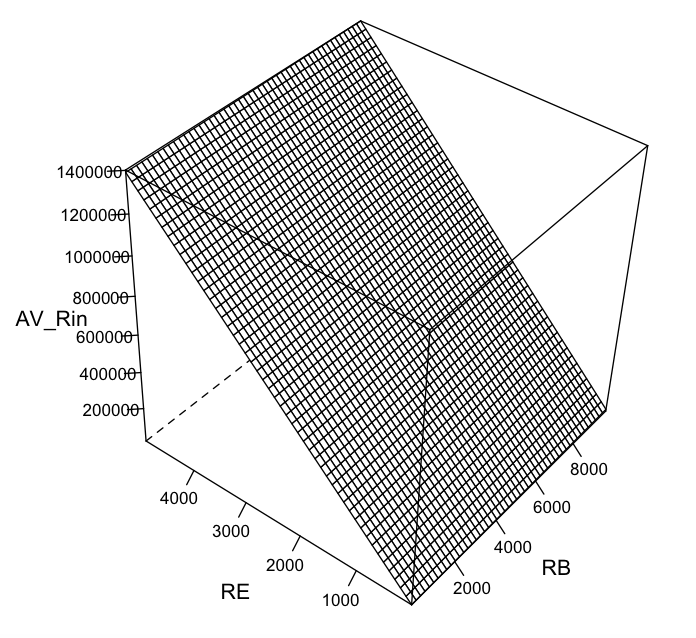}
\includegraphics[width=8cm]{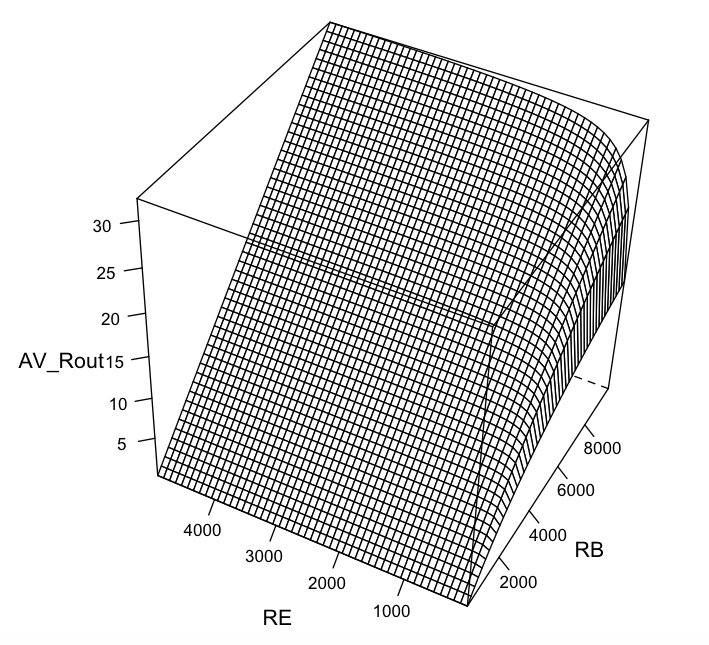} \\
(c)   \hspace{7.5cm}  (d)
\caption{Properties of the common collector amplifier for a prototypic PNP transistor: (a) average voltage gain; (b) THD;
(c) average input resistance; and (d) average output resistance.   $V_a = -50 V$ and $s = 5.0 V^{-1}$.}
\label{fig:RBRE_PNP}}
\end{figure*}

\begin{figure*}[h!]
\centering{
\includegraphics[width=8cm]{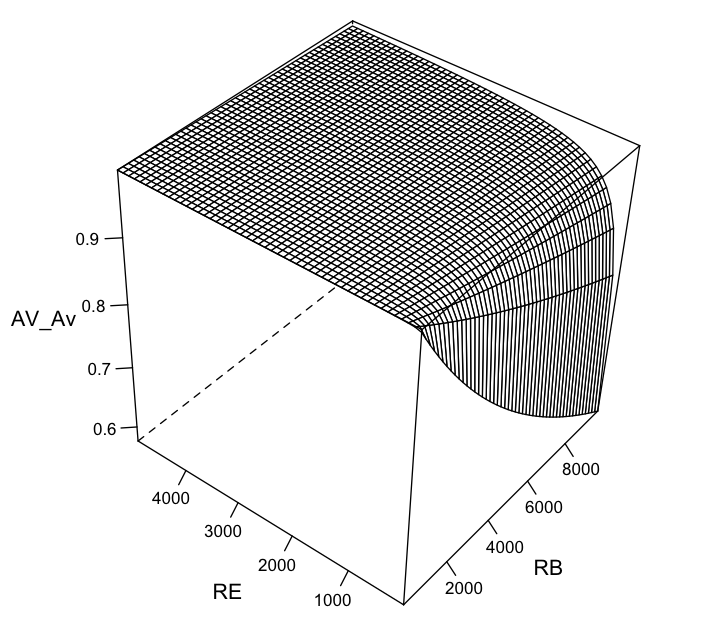}
\includegraphics[width=8cm]{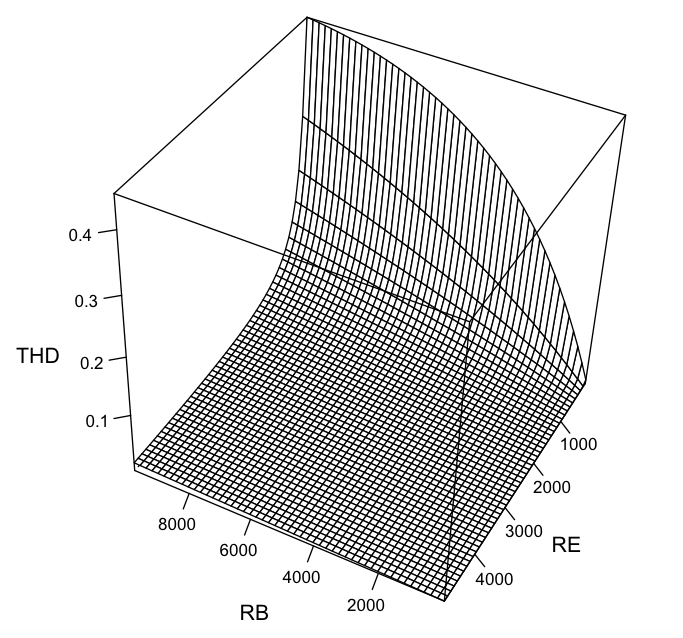}   \\
(a)     \hspace{7.5cm}     (b)   \\
\includegraphics[width=8cm]{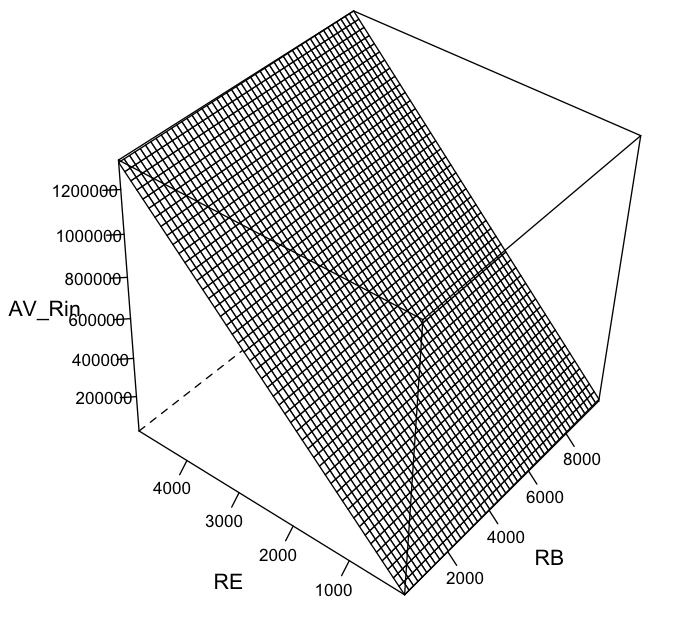}
\includegraphics[width=8cm]{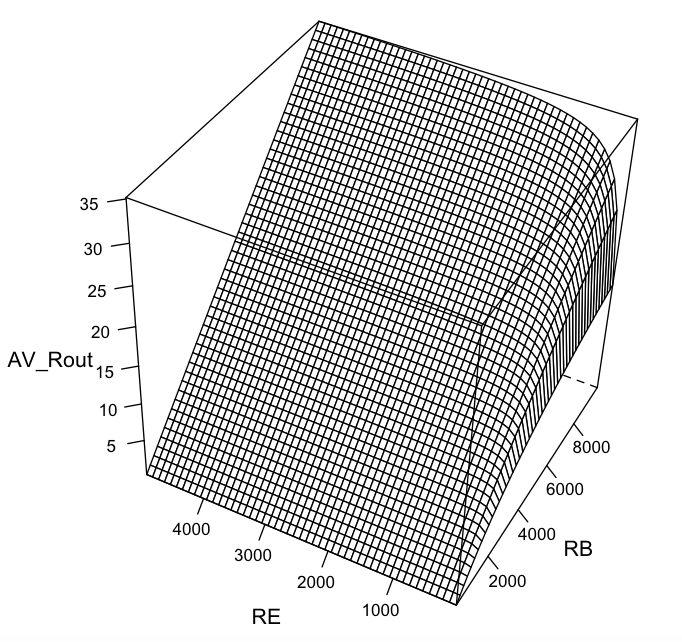} \\
(c)   \hspace{7.5cm}  (d)
\caption{Properties of the common collector amplifier for a prototypic NPN transistor: (a) average voltage gain; (b) THD;
(c) average input resistance; and (d) average output resistance. $V_a = -150 V$ and $s = 2.5 V^{-1}$.}
\label{fig:RBRE_NPN}}
\end{figure*}

Figure~\ref{fig:RBRE_PNP} shows, respectively, the average voltage gain (a), the THD (b), the average input resistance (c) 
and the average output resistance (d) in terms of $R_B$ and $R_E$ obtained with respect to the PNP prototypical transistor.  Figure~\ref{fig:RBRE_NPN} illustrates the same respective circuit properties as obtained for the NPN prototypical transistor.

We have that the respective surfaces obtained for the PNP and NPN small signal transistors have very similar shapes.  In the 
case of the average gain, even the values are very similar.  However, rather distinct distortions (THD) were obtained, with the NPN
case yielding about half THD intensities.  This suggests that NPN transistors can be used in cases where linearity is more critical.
Similar, but not identical, average input and output resistances were obtained for the NPN and PNP cases.

Regarding the effects of $R_B$ and $R_E$ choices on the circuit properties, we have that the average voltage gain is very
stable and close to unit in the region where $R_E$ is larger and $R_B$ is smaller.  The average gain tends to decrease more
substantially for larger values of $R_B$, and a drastic gain variation is observed when $R_E$ approaches $100\Omega$.
The latter effect shows that the reduction of the negative feedback level undermines the unit gain associated with the common
collector amplifier.  

The THD behavior with $R_B$ and $R_E$, it takes small valuer for large values of $R_E$ almost irrespectively with $R_B$,
but undergoes a dramatic increase for smaller values of $R_E$.   This effect again shows the importance of adopting relatively high
values of $R_E$ in order to enhance the negative feedback implemente by this resistance.  

The average input resistance varies in almost perfectly linear way with $R_E$, irrespectively of $R_B$, achieving 
$\approx 1.4 M \Omega$ for the PNP transistor and $1.2 M \Omega$ for the NPN transistor.  This is so because the latter type 
of transistors have smaller output resistance $R_o \approx 1/(s I_B)$ than a PNP counterpart with the same average current 
gain $\approx -s V_a$.

Now, it is interesting to observe the average input and output resistances obtained for the region of smallest THD, namely
the configurations near $(R_B = 1000 \Omega, R_E =5000 \Omega)$.  At this point, the average input resistance is at its
highest, and the average output resistance is at its lowest.  In addition, the average voltage gain is also observed to be at
the nearest value to unit. So, this configuration seems to be ideal when the common collector amplifier is used as a buffer,
be it employing PNP or NPN transistors.  

As expected, very small average output resistances are obtained for both PNP and NPN cases, with values slightly larger
observed for the NPN transistor.  The average output resistance increases almost linearly with $R_B$ for most values of
$R_E$, but undergoes an abrupt transition for the smallest values of this resistance.

\begin{figure*}[h!]
\centering{
\includegraphics[width=8cm]{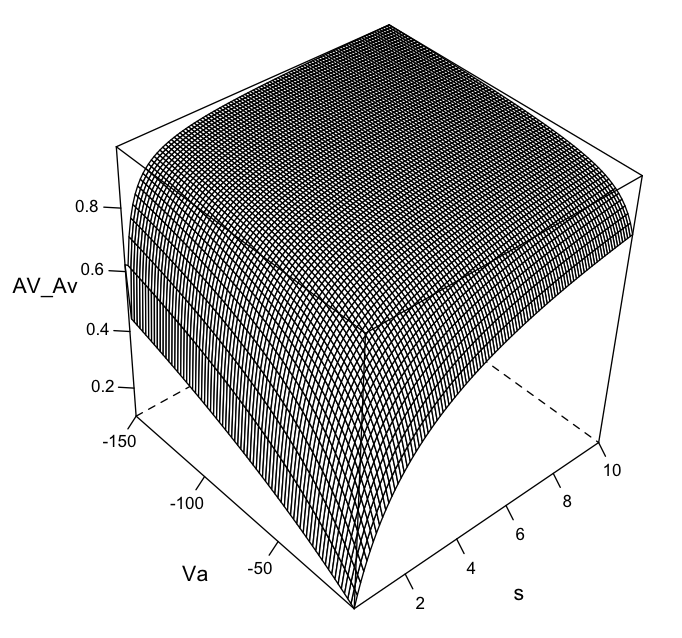}
\includegraphics[width=8cm]{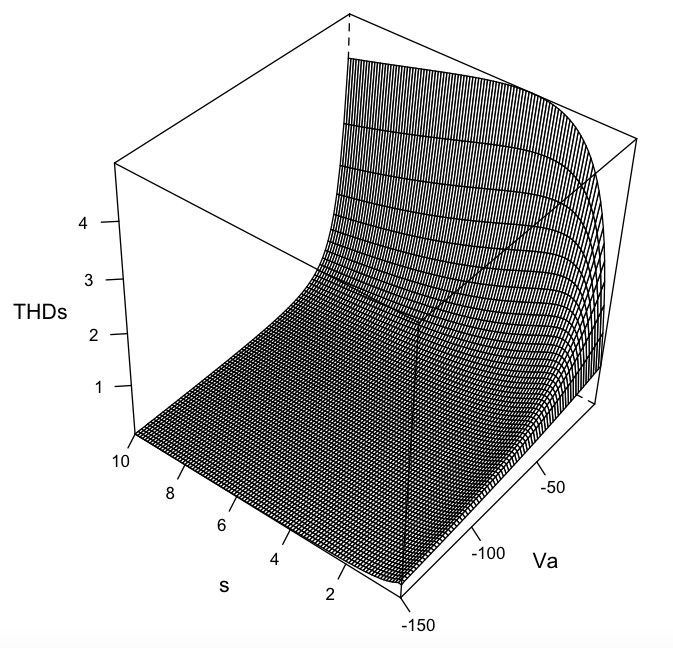}   \\
(a)     \hspace{7.5cm}     (b)   \\
\includegraphics[width=8cm]{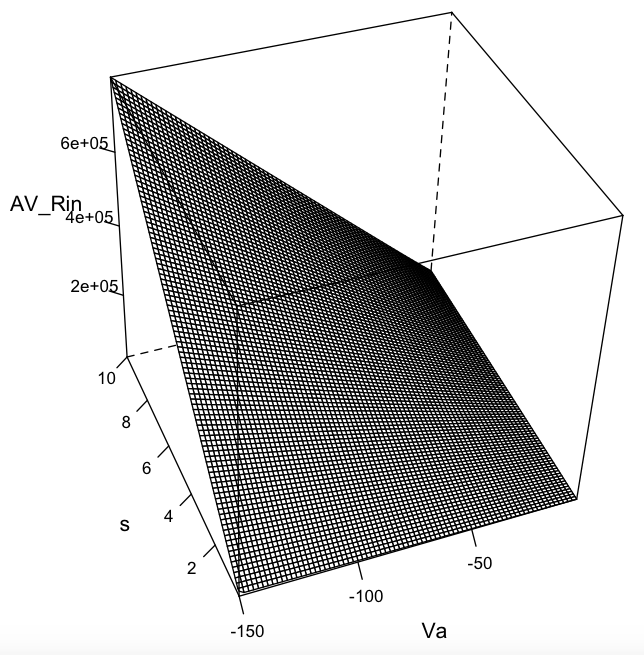}
\includegraphics[width=8cm]{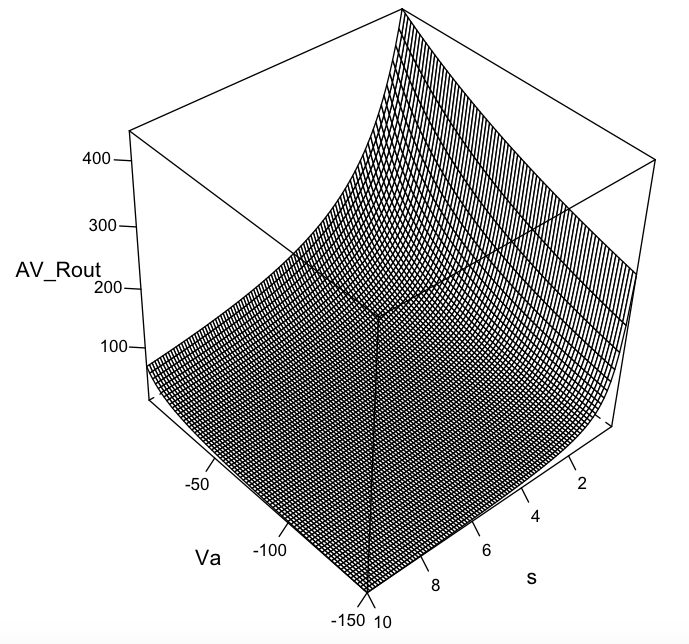} \\
(c)   \hspace{7.5cm}  (d)
\caption{Properties of the common collector amplifier for a prototypic NPN transistor: (a) average voltage gain; (b) THD;
(c) average input resistance; and (d) average output resistance.  $R_B = 10 k \Omega$ and $R_E = 1 k \Omega$ and 
$R_L = R_e = 500 \Omega$.}
\label{fig:sVa}}
\end{figure*}

Now, in order to have a better idea of how the transistor parameters $V_a$ and $s$ affect the common collector voltage
gain and input and output resistances, we obtain these properties for a wide range of these parameters while fixating
$R_B = 10 k \Omega$ and $R_E = 1 k \Omega$ and $R_L = R_e = 500 \Omega$.  Figure~\ref{fig:sVa} depicts the respectively 
obtained average voltage gain (a), the THD (b), as well as the average input resistance (c) and average output resistance (d).

First we analyze the average voltage gain, shown in Figure~\ref{fig:sVa}(a).  This gain is immediately verified to take its
largest values (nearer to 1) in the region of the Early space characterized by large $V_a$ magnitude and large values of
$s$.  So, gain nearer to one is obtained for the largest average current gain $\langle \beta \rangle \approx -s V_a$.
However, silicon small signal transistors have verified~\cite{costaearly:2018} to have approximately
$130 \leq \langle \beta \rangle \approx 400$, achieving an average gain value of 250.  This defines an approximate curve
$s = -250/V_a$ in the Early space, representing the band typically occuppied by silicon small signal transistors.  It can
be verified that, for the chosen values of o$R_B$ and $R_E$, the gain along this curve departs significantly from the unit.
As discussed above, better perfomance can be obtained for larger values of $R_E$.

The THD obtained in terms of $(V_a, s)$ is shown in Figure~\ref{fig:sVa}(b).  It follows from this result that best linearity
is achieved for large values of both $V_a$ and $s$.  However, for the region typically occupied by silicon small signal
transistors.   Because of the asymmetry of this surface, better THD is obtained for NPN devices, which tend to have
larger $V_a$ magnitude and $s$ values, than for PNP transistors.

The average input resistance shows a definite increase with both $V_a$ magnitude and $s$ values, reaching its peak
at $(V_a = -150, s= 10)$.  However, for the typical Early parameter values characterizing silicon small signal
transistors, the average input resistance will be much smaller than at its peak, achieving typically $200 k \Omega$, be
it for NPN or NPN devices.

The smallest average output resistance, illustrated in Figure~\ref{fig:sVa}(d), is again obtained for $(V_a = -150, s= 10)$,
which corresponds to the maximum average current gain.  Sub-optimal values are verified along the band occupied
typically by NPN and PNP silicon small signal transistors.

All in all, the mapping of the four important circuit properties in terms of the Early parameters of the transistors confirmed
that the best properties are obtained for the largest average current gain $\approx 1500$, as this is traded-off for linearity and other
benefits.  However, real-world small signal transistors have much smaller average current gains, hence limited performances
that, as in the case of THD, varies with the choice of specific transistor parameters.  This is particularly interesting because
it was thanks to the consideration of the transistor linearities allowed by the adopted Early modeling approach that this
important effect on the linearity has been observed, while going unnoticed otherwise.

\section{Concluding Remarks}

We have reported the application of the recently introduced Early modeling approach to the analysis of the common collector
amplifier, also known as emitter follower.  The main distinctive features of this approach are: (i) the whole approach is very
simple and involves small computational resources; (ii) it allowed the non-linearities of the transistors implied by the Early effect 
to be accurately taken into account; and (iii) it allowed mathematical expressions to be obtained describing the involved
voltages and currents.  

The Early equivalent circuit of a transistor was used to obtain to transform the common collector amplifier into a mathematically
approachable representation, from which expressions describing the circuit operation in terms of currents and voltages were
obtained.  These equations were then used to obtain the average voltage gain and the THD, as well as the average input
and output resistances.  Prototypes of PNP and NPN small signal silicon transistors obtained previously~\cite{costaearly:2018}
were used to discuss the effect of the choice of the resistances $R_B$ and $R_E$ on the circuit performance.   This study
was complemented by obtaining and discussing the effect of the transistor Early parameters $V_a$ and $s$ on the common
collector operation.

The obtained results confirmed the trade-off of current gain for improved linearity, larger input resistance, and lower output
resistance, as well as near unit voltage gain.  However, it was also shown that twice as much THD can be obtained for the
same circuit configuration (i.e.~same values of $R_B$ and $R_E$) when using PNP devices instead of NPN counterparts
with the same average current gain.   This important result was only possible because of the Early modeling approach ability
to take into account the non-linearity of small signal transistors implied by the Early effect and corresponding to the
converging beta-indexed characteristic isolines.  

The reported procedure and results pave the way to a number of related further developments, including the analysis of the
dispersion of the four considered measurements (i.e.~average voltage gain and THD, as well as average input and output
resistances).  It would also be interesting to extend the reported analysis to higher power and higher frequency transistors, as
well as other technologies such as FET and MOS.  Other circuit configurations such as the differential amplifier can also be
approached by using the reported methodology.

\vspace{0.7cm}
\textbf{Acknowledgments.}

Luciano da F. Costa
thanks CNPq (grant no.~307333/2013-2) for sponsorship. This work has benefited from
FAPESP grants 11/50761-2 and 2015/22308-2.
\vspace{1cm}

\bibliography{mybib}
\bibliographystyle{unsrt}
\end{document}